
\magnification 1200
\baselineskip=6truemm


\def\npb#1#2#3{{\it Nucl. Phys.} {\bf B#1} (19#2) #3}
\def\plb#1#2#3{{\it Phys. Lett.} {\bf #1B} (19#2) #3}
\def\prd#1#2#3{{\it Phys. Rev.} {\bf D#1} (19#2) #3}


\def\dbar{\overline d}
\def\ubar{\overline u}
\def\nubar{\overline\nu}
\def\small#1{{\scriptscriptstyle #1}}
\def\Scl{{\cal L}}
\def\Scm{{\cal M}}
\def\sla#1{\raise.15ex\hbox{$/$}\kern-.57em #1}
\def\geapprox{\lower.6ex
              \hbox{\kern+.2em$\buildrel >\over\sim$\kern+.2em}}
\def\leapprox{\lower.6ex
              \hbox{\kern+.2em$\buildrel <\over\sim$\kern+.2em}}


\def\ginzburg{1}
\def\rizhew{2}
\def\buchwyler{3}
\def\belgeng{4}
\def\lep{5}
\def\rizzo{6}
\def\uatwo{7}
\def\ruckl{8}
\def\heraone{9}
\def\heratwo{10}
\def\hewett{11}

\line{\hfil January, 1993}
\line{\hfil UdeM-LPN-TH-132}
\null\vskip 3truecm
\centerline{\bf Leptoquarks at $e\gamma$ Colliders}
\vskip 2truecm
\centerline{H\'el\`ene Nadeau and David London}
\vskip .15in
\centerline{\it Laboratoire de Physique Nucl\'eaire, Universit\'e de
Montr\'eal}
\centerline{\it C.P. 6128, Montr\'eal, Qu\'ebec, CANADA, H3C 3J7.}
\vskip 1truecm
\centerline{\it ABSTRACT}
\noindent
We consider single production of leptoquarks (LQ's) at $e\gamma$ colliders,
for  two values of the centre-of-mass energy, $\sqrt{s}=500$ GeV and 1 TeV.
LQ's with masses essentially up to the kinematic limit can be seen, even
for couplings as weak as $O(10^{-3})$-$O(10^{-2})\alpha_{em}$. It is
possible to detect LQ's of mass greater than $\sqrt{s}$ by looking for
signals of virtual LQ production in $e\gamma\to e q{\overline q}$.
\vfill\eject


One of the colliders planned for the coming decade is the Next Linear
Collider (NLC)\footnote*{By NLC, we mean any of the linear colliders which
are now being discussed (JLC, VLEPP, NLC/TLC, CLIC, DESY-THD, TESLA,
...).},  a linear $e^+e^-$ collider with a centre-of-mass energy in the
range of about 500 GeV-1 TeV. This will provide a very clean environment in
which to look for and study new physics not found in the standard model
(SM). Some time ago it was noted [\ginzburg] that these machines could be
transformed into $e\gamma$ or $\gamma\gamma$ colliders of roughly the same
energy and luminosity by backscattering laser light from one or both
electron beams. Work has already begun which looks at the prospects for the
observation of new physics in $e\gamma$ and $\gamma\gamma$ collisions.

In this paper we consider the $e\gamma$ option and investigate the
possibilities for detecting leptoquarks (LQ's) at such a machine.
Leptoquarks appear in a large number of extensions of the SM such as grand
unified theories (GUT's), technicolor, and composite models. In general,
these particles can have either spin 0 or spin 1, but here we consider only
scalar (spin-0) LQ's. There are four possible leptoquark charges:
$Q_{em}=-1/3$, $-2/3$, $-4/3$ and $-5/3$. These correspond to the decays of
the LQ into $e^-u$, $e^-\dbar$, $e^-d$ and $e^-\ubar$, respectively. For
the charge $-$1/3 and $-$2/3 LQ's, there may be other decay modes, such as
$\nu d$ or $\nubar d$ and $\nu\ubar$ or $\nubar\,\ubar$.

The leptoquarks most commonly considered in the literature are those with
$Q_{em}=-1/3$. These LQ's appear, for example, in $E_6$
superstring-inspired grand unified theories [\rizhew]. If these leptoquarks
are light (i.e.~observable at collider energies), then a large fraction of
the parameter space of their couplings has already been ruled out
[\buchwyler]. Couplings of the LQ and the lepton-quark pair which are
flavour-violating will typically lead to low-energy flavour changing
neutral currents (FCNC's), and are hence constrained to be small. Thus, the
only allowed sizable LQ couplings are those which are flavour-diagonal.
This implies that only leptoquarks which couple to first-generation
particles can be usefully studied at an $e\gamma$ collider, since the LQ's
necessarily interact with the electron in these collisions. In addition, in
order to evade constraints from rare pion decays such as $\pi^+\to
e^+\nu_e$, the couplings of such LQ's must be chiral. Finally, a comparison
of the Fermi constants extracted from $\mu$- and $\beta$-decay
(quark-lepton universality) requires that the couplings of the left-handed
LQ be at most about 10\% of electromagnetic strength. This is due to the
fact that the LQ contribution to these processes can interfere with that of
the standard model. There is no analogous constraint on the right-handed LQ
coupling.

We must reiterate that the above constraints apply only to the
$Q_{em}=-1/3$ leptoquarks. They do {\it not} necessarily apply to LQ's of
other charges. For example, many of the limits are derived using the fact
that the charge $-1/3$ LQ couples both to $e^-u$ and $\nu_e d$. As will be
seen below, the other LQ's couple either to $e^-$ and (anti)quark or
$\nu_e$ and (anti)quark, but not both. Thus there is no constraint from
$\pi^+\to e^+\nu_e$, for example. Nevertheless, in this paper we will
follow the conventions used in the literature -- we will assume that the
couplings of all leptoquarks are flavour-diagonal, and we will present our
results in terms of the chiral couplings of all LQ's. However, the reader
should be aware that the couplings of LQ's with $Q_{em}\ne -1/3$ are not
necessarily required phenomenologically to follow these patterns
[\belgeng].

It might be argued that, in any event, the leptoquarks most likely to exist
are those with charge $-1/3$ since they are the best-motivated
theoretically. We do not subscribe to this. In general, the $E_6$ models
which predict the existence of such LQ's also predict that these particles
acquire a mass of the order of the scale at which $E_6$ is broken, i.e.~the
GUT scale, so that there are no light LQ's. It may be possible to concoct
variants in which additional adhoc discrete symmetries are introduced which
keep the LQ's light, but these are not particularly well-motivated.
Instead, we take---and we encourage the experimentalists to take---a more
model-independent point of view, and consider the production of leptoquarks
of all charges, both real and virtual, at an $e\gamma$ collider.

Leptoquark production at $e^+e^-$, $p{\overline p}$ and $ep$ colliders has
also been discussed in the literature. Results from LEP imply that the LQ
mass is greater than $M_Z/2$ [\lep]. This could be extended to $\sim 80$
GeV by looking for single leptoquarks [\rizzo]. The limits from UA2 depend
somewhat on the coupling strength, but give a lower bound on LQ masses of
about 70 GeV [\uatwo]. The $ep$ collider HERA is ideally suited to the
search for leptoquarks, and it is expected that, for a large range of
couplings, LQ's with masses up to the kinematic limit can be seen
[\ruckl,\heraone,\heratwo]. As we will show, $e\gamma$ colliders are also
very useful tools in searching for light leptoquarks, since these can be
produced singly in such experiments. Single leptoquark production in
$e\gamma$ collisions has been considered by Hewett and Pakvasa [\hewett] in
the context of $e^+e^-$ colliders. We will apply their results specifically
to $e\gamma$ colliders and extend them to include the possibility of single
virtual leptoquark production.

Buchm\"uller et.\ al.\ [\ruckl] have written down the most general
$SU(3)\times SU(2)\times U(1)$ invariant scalar leptoquark couplings which
satisfy baryon and lepton number conservation. Defining the fermion number
$F=3B+L$, where $B$ is baryon number and $L$ is lepton number, it is
possible to have leptoquarks whose couplings obey $\vert\Delta F\vert=0$ or
2 (we follow the notation of Ref.~\rizzo):
$$\eqalign{
\Scl_{\vert\Delta F\vert=2} = & ~g_{1\small{L}}
{\overline{q^c_\small{L}}}i\tau_2 l_\small{L} S_1 + g_{1\small{R}}
{\overline{u_\small{R}^c}}e_\small{R} S_1^\prime  + {\tilde g}_{1\small{R}}
{\overline{d_\small{R}^c}}e_\small{R} {\tilde S}_1 + g_{3\small{L}}
{\overline{q_\small{L}^c}}i\tau_2 \tau^i l_\small{L} S_3^i, \cr
\Scl_{\vert\Delta F\vert=0} = & ~h_{2\small{L}} {\overline u}_\small{R}
l_\small{L} R_2 + h_{2\small{R}} {\overline q}_\small{L} i \tau_2 e_\small{R}
R_2^\prime  + {\tilde h}_{2\small{L}} {\overline d}_\small{R} l_\small{L}
{\tilde R}_2, \cr}\eqno(1)
$$
where $q_\small{L}$ and $l_\small{L}$ are the standard left-handed
$SU(2)_L$-doublets of  quarks and leptons, respectively, and
$\psi^c=C{\overline\psi}^\small{T}$ is a charge-conjugated fermion field. The
leptoquarks $S_1$, $S_1^\prime$ and ${\tilde S}_1$ are $SU(2)_L$ singlets,
$R_2$, $R_2^\prime$ and ${\tilde R}_2$ are $SU(2)_L$ doublets, and $S_3$ is
an $SU(2)_L$ triplet.

Only those LQ's which couple to electrons can be produced, either as real
or virtual particles, in $e\gamma$ collisions. There are eight such types,
distinguished by the (anti)quark which is produced along with the (real or
virtual) leptoquark in the final state, as well as the chirality of the LQ
coupling (chirality here is defined as the helicity of the electron). These
are listed in Table 1.

$$
\vbox{\offinterlineskip
\halign{&\vrule#&
   \strut\quad#\hfil\quad\cr
\noalign{\hrule}
height2pt&\omit&&\omit&&\omit&&\omit&\cr
& Process && $Q_{\small{S}}$ && Chirality && Origin &\cr
height2pt&\omit&&\omit&&\omit&&\omit&\cr
\noalign{\hrule}
height2pt&\omit&&\omit&&\omit&&\omit&\cr
& $e^-\gamma\to S \ubar$ && $-1/3$ && L.H. && $S_1$, $S_3$ &\cr
& $e^-\gamma\to S \ubar$ && $-1/3$ && R.H. && $S_1^\prime$ &\cr
& $e^-\gamma\to S \dbar$ && $-4/3$ && L.H. && $S_3$ &\cr
& $e^-\gamma\to S \dbar$ && $-4/3$ && R.H. && ${\tilde S}_1$ &\cr
& $e^-\gamma\to S u$ && $-5/3$ && L.H. && $R_2$ &\cr
& $e^-\gamma\to S u$ && $-5/3$ && R.H. && $R_2^\prime$ &\cr
& $e^-\gamma\to S d$ && $-2/3$ && L.H. && ${\tilde R}_2$ &\cr
& $e^-\gamma\to S d$ && $-2/3$ && R.H. && $R_2^\prime$ &\cr
height2pt&\omit&&\omit&&\omit&&\omit&\cr
\noalign{\hrule}}}
$$
\noindent
Table 1: The eight types of leptoquark ($S$) which can be produced in
$e\gamma$ collisions. The $S$ can be real or virtual. The `chirality'
refers to the helicity of the $e^-$ -- either left-handed (L.H.) or
right-handed (R.H.).
\vskip3truemm

Let us first turn to real leptoquark production. The diagrams which are
responsible for single LQ production in $e\gamma$ collisions are shown in
Fig.~1. For the production of real leptoquarks, the chirality of the
coupling is irrelevant. The cross sections which we will calculate are
equal for left-handed and right-handed leptoquarks. (Of course, if a
leptoquark were detected, electron polarization could be used to ascertain
its origin.) In this case, there are essentially four types of LQ,
distinguished by their electromagnetic charge. These four types of LQ can
be compared by simply keeping the charge of the leptoquark,
$Q_{\small{S}}$, as a parameter in the calculations of the diagrams of
Fig.~1.

Replacing the couplings $g_{1\small{L}}$, ..., ${\tilde h}_{2\small{L}}$ of
Eq.~1 by a generic leptoquark Yukawa coupling $g$, we conventionally
parametrize $g$ by scaling it to electromagnetic strength, $g^2/4\pi =
k\alpha_{em}$, and allow $k$ to vary. With this notation, the differential
cross section for the diagrams of Fig.~1 is then found to be
$$\eqalign{
{d\sigma\over du} = & -{\pi k\alpha^2_{em} \over 2 s^2} \left\{
-{u\over s} - 2 Q_{\small{S}}^2 {u(u+s-M_{\small{S}}^2)^2\over s(u+s)^2}
- 2 Q_{\small{S}} {u(u+s-M_{\small{S}}^2)\over s(u+s)} \right. \cr
&~~~~~
- (1+Q_{\small{S}})^2 \left[ {s\over u} +
2{(u-M_{\small{S}}^2)(u+s-M_{\small{S}}^2)\over s u} \right]
+ 2 (1+Q_{\small{S}}){u-M_{\small{S}}^2\over s} \cr
&~~~~~ \left. +4 Q_{\small{S}} (1+Q_{\small{S}})
{(u+s-M_{\small{S}}^2)(s/2+u-M_{\small{S}}^2)\over s(u+s)} \right\}, \cr}
\eqno(2)$$
in which $u=-s\beta{1\over 2}(1+\cos\theta)$, with
$\beta=1-M_{\small{S}}^2/s$. (We note that this expression agrees with that
of Ref.~\hewett\ for $Q_{\small{S}}=-1/3$, with the replacement $k \to 2k$.
This change in $k$ is due to a difference of $\sqrt{2}$ between us and the
authors of Ref.~\hewett\ in the definition of the LQ coupling of Eq.~1.)

In fact, Eq.~2 can be integrated analytically. Including a $p_{\small{T}}$
cut on the associated jet of $p_{\small{T}_{jet}} > 10$ GeV, we present the
results in Fig.~2, for two values of $\sqrt{s}$, 500 GeV and 1 TeV, for
each of the four LQ charges, with $k=1$. The LQ's with charge $-5/3$ and
$-1/3$, which couple the electron to ${\overline u}$- or $u$-quarks, have
larger cross sections than the charge $-4/3$ and $-2/3$ LQ's, which couple
to ${\overline d}$- or $d$-quarks. This is due to the difference of the
quark charges, which is important in Fig.~1c. The charge of the LQ's
themselves play a role (see Fig.~1b), as can be seen by the fact that the
charge $-5/3$ LQ has a larger cross section than the charge $-1/3$ LQ, and
similarly for the other two charges. This effect is of course largest for
light leptoquarks, where the diagram in Fig.~1b is particularly important.
The expected luminosity at the NLC is 10 ${\rm fb}^{-1}$ at 500 GeV, and 60
${\rm fb}^{-1}$ at 1 TeV. Thus, depending on the charge of the leptoquark,
one expects somewhere between $\sim 200$ and $\sim 6000$ events for $k=1$,
for the LQ mass up to essentially the kinematic limit.

Except for the left-handed charge $-1/3$ LQ's $S_1$ and $S_3$, all the
leptoquarks in Table 1 decay exclusively to $eq$ or $e{\overline q}$, so
the final state will be an electron and 2 jets. $S_1$ and $S_3$ can also
decay to $\nu d$, which gives a final state of 2 jets and missing energy.
Both of these final states, $ejj$ and $\sla{p}_\small{T} jj$, can be
produced via standard model processes, namely $e\gamma\to eZ\to ejj$ and
$e\gamma\to \nu W\to \sla{p}_\small{T} jj$. However, these will cause no
problems. First of all, the leptoquark signal includes a sharp invariant
mass peak in $M_{ej}$ or $M_{\sla{p}_\small{T} j}$ at $M_{\small{S}}$,
which is not present in the SM decays. Rather, the SM processes will
produce mass peaks in $M_{jj}$ at $M_Z$ or $M_W$ (which are absent in the
case of leptoquark production). Also, in $W$- or $Z$-decay, the jets will
tend to be colinear, while the jets in leptoquark decay will be widely
separated. Thus, a simple cut on the angle between the jets should
eliminate essentially all the background, and so the discovery limit is
simply set by the signal rate. This means that LQ's whose coupling strength
is considerably weaker than electromagnetic can be detected. This is made
more quantitative in Fig.~3, where we plot the discovery limits (defined as
25 events) as a function of $M_{\small{S}}$ and $k$ for each of the 4
charges of leptoquark. From this figure it is seen that leptoquarks with
$k$ as small as $O(10^{-3})$-$O(10^{-2})$ can be detected for most of the
kinematically allowed mass range.

One can try to go beyond the kinematic limit by considering virtual single
leptoquark production, as shown in Figs.~4a-c. The problem here is that the
process of interest is $e\gamma\to e q{\overline q}$, which has an enormous
standard model background (Figs.~4d,e). It is therefore necessary to
calculate the cross section for this process including all the diagrams of
Fig.~4, and to look for a set of cuts which reduces the SM contribution to
a level where the presence of leptoquarks is observable. This is what we
have done. We will see that it is possible to detect LQ's whose mass is
greater than the kinematic limit, but to go well beyond this limit requires
that $k>1$, i.e.~that the coupling be stronger than that of the
electromagnetic interaction.

The cross section is calculated for all eight types of LQ shown in Table 1.
We will keep track of the charge of the leptoquark as in the previous
calculation of real leptoquark production. As to the LQ chirality, it is
important only in the interference of the leptoquark and SM diagrams -- it
indicates which of the left- and right-handed couplings of the photon/$Z$
is selected in these terms. For example, the right-handed charge $-1/3$ LQ
couples to right-handed electrons and right-handed $u$-quarks -- thus, only
the SM diagrams with right-handed couplings will interfere with the LQ
diagrams.

There is one technical point regarding the calculation. The momentum
assignments of the initial and final state particles are shown in Fig.~4a.
As is evident from this figure, the momenta of the final quark and
antiquark depend on whether the leptoquark has fermion number $F=0$ or 2.
The LQ's of charge $-2/3$ and $-5/3$ have fermion number 0, while those
with $Q_{em}=-1/3$ and $-4/3$ have $F=2$. If $F=0$, the quark has momentum
$p_4$ and the antiquark $p_6$, while for $F=2$, these are switched. As far
as the leptoquark diagrams are concerned, this is irrelevant, but it is
important for the SM contribution. We take this point into account in the
expressions below.

The contribution to the total cross section for $e\gamma\to e q{\overline
q}$ which is due solely to leptoquarks (Figs.~4a-c) is calculated to be
$$\eqalign{
{1\over 4}\sum_{\rm spins} \vert\Scm_{\small{L}\small{Q}}\vert^2 = &
{}~128\pi^3 k^2 \alpha^3_{em}
{p_5\cdot p_6\over \left[2 p_5\cdot p_6 - M_{\small{S}}^2\right]^2}
\left\{ {p_1\cdot p_4\over s} + 8 Q_{\small{S}}^2
{p_1\cdot p_4 \, (p_2\cdot p_4)^2\over
		s \left[2 p_2\cdot p_4 + M_{\small{S}}^2\right]^2} \right. \cr
&~~~~~ + {1\over 4} (1 + Q_{\small{S}})^2 {s\over p_1\cdot p_4}
\left[ 1 + 8 \left( p_2\cdot p_4 - {s\over 2} \right)
{p_2\cdot p_4\over s^2}\right] \cr
&~~~~~ + 4 Q_{\small{S}} { p_1\cdot p_4 \, p_2\cdot p_4 \over
			s \left[2 p_2\cdot p_4 + M_{\small{S}}^2\right]}
- (1+Q_{\small{S}}) \left[ 1 - 2 {p_2\cdot p_4\over s} \right] \cr
&~~~~~ \left. - 2 Q_{\small{S}} (1+Q_{\small{S}})
{ p_2\cdot p_4 \over \left[ 2 p_2\cdot p_4 + M_{\small{S}}^2\right]}
\left( 1 - 4 {p_2\cdot p_4\over s} \right) \right\}, \cr} \eqno(3)
$$
where $M_{\small{S}}$ is the mass of the leptoquark, $Q_{\small{S}}$ is its
charge, and the momenta $p_1$-$p_6$ are defined in Fig.~4a.

The SM diagrams (Figs.~4d,e) include the contributions of both the photon
and the $Z$, which we will denote $Z_0$ and $Z_1$, respectively. Writing
the $Z_i f{\overline f}$ couplings as
$$
-ie(v_i^f\gamma^\mu-a_i^f\gamma^\mu\gamma_5),\eqno(4)
$$
it is useful to define
$$\eqalign{
B_{ij} \equiv & ~(v_i^q v_j^q + a_i^q a_j^q ) (v_i^e v_j^e + a_i^e a_j^e), \cr
E_{ij} \equiv & ~(v_i^q a_j^q + a_i^q v_j^q ) (v_i^e a_j^e + a_i^e v_j^e), \cr
}\eqno(5)$$
where $q=d,u$ and $i,j=0,1$. We also define
$$
A_{ij} \equiv
{ (p_3^2 - M_i^2 ) (p_3^2 - M_j^2 ) + (\Gamma_i M_i) (\Gamma_j M_j) \over
\left[ (p_3^2 - M_i^2 )^2 + (\Gamma_i M_i)^2 \right]
\left[ (p_3^2 - M_j^2 )^2 + (\Gamma_j M_j)^2 \right] }~,\eqno(6)
$$
where $i,j=0,1$ and $p_3^2=(p_4+p_6)^2$. The SM contribution to the cross
section $e \gamma \to e q {\overline q}$ is then
$$\eqalign{
{1\over 4}\sum_{\rm spins} & \vert\Scm_{\small{S}\small{M}}\vert^2 =
{}~{1024 \pi^3 \alpha^3_{em} \over s} \sum_{ij} A_{ij} \Bigl\{ \cr
& B_{ij} {1\over p_1\cdot p_5 }
\left[ s\, \epsilon_k\cdot p_4 \, \epsilon_k\cdot p_6 \,
p_1\cdot p_5 + s \, p_1\cdot p_5 \, p_4\cdot p_6 + 2 \, p_1\cdot p_4 \,
p_1\cdot p_6 \, p_2\cdot p_5 \right] \cr
& + \left( B_{ij} - (-1)^{(F/2)} E_{ij} \right) \Bigl[ p_1\cdot p_4
\, p_5\cdot p_6 \cr
& \qquad + {1\over 2 ( p_1\cdot p_5 )} \bigl(
s \, \epsilon_k\cdot p_4 \, \epsilon_k\cdot p_5 \, p_6 \cdot (p_5-p_1)
- s \, p_1\cdot p_6 \, p_4\cdot p_5 \cr
& \qquad\qquad\qquad\qquad\qquad\qquad
- 2 \, p_1\cdot p_4 \, p_1\cdot p_5 \, p_2\cdot p_6 \bigr) \cr
& \qquad + {s\over 2 ( p_1\cdot p_5 )^2 } \bigl(
\left(\epsilon_k\cdot p_5\right)^2 p_2\cdot p_4 \, p_6\cdot (p_5 - p_1)
+ \epsilon_k\cdot p_5 \, \epsilon_k\cdot p_6 \, p_1\cdot p_5 \, p_2\cdot p_4
\cr
& \qquad\qquad\qquad\qquad\qquad\qquad + p_1\cdot p_5 \, p_1\cdot p_6 \,
p_2\cdot p_4  \bigr) \Bigr] \cr
& + \left( B_{ij} + (-1)^{(F/2)} E_{ij} \right) \Bigl[ p_1\cdot p_6
\, p_4\cdot p_5 \cr
& \qquad + {1\over 2 ( p_1\cdot p_5 )} \bigl(
s \, \epsilon_k\cdot p_5 \, \epsilon_k\cdot p_6 \, p_4 \cdot (p_5-p_1)
- s \, p_1\cdot p_4 \, p_5\cdot p_6 \cr
& \qquad\qquad\qquad\qquad\qquad\qquad
- 2 \, p_1\cdot p_6 \, p_1\cdot p_5 \, p_2\cdot p_4 \bigr) \cr
& \qquad + {s\over 2 ( p_1\cdot p_5 )^2 } \bigl(
\left(\epsilon_k\cdot p_5\right)^2 p_2\cdot p_6 \, p_4\cdot (p_5 - p_1)
+ \epsilon_k\cdot p_4 \, \epsilon_k\cdot p_5 \, p_1\cdot p_5 \, p_2\cdot p_6
\cr
& \qquad\qquad\qquad\qquad\qquad\qquad + p_1\cdot p_4 \, p_1\cdot p_5 \,
p_2\cdot p_6 \bigr) \Bigr]
\Bigr\}, \cr}\eqno(7)
$$
where $F$ is the fermion number of the leptoquark (as discussed above, the
fermion number of the leptoquark defines the momenta of the quark and
antiquark). The $\epsilon_k$ are the two polarizations of the initial
photon. In the above expression, wherever two $\epsilon_k$'s appear, a sum
over $k=1,2$ is implicit.

Finally, the interference terms between the LQ and SM diagrams remain to be
calculated. As noted earlier, either the left- or right-handed SM couplings
to the electron and quark contribute in this interference, but not both. It
is the leptoquark chirality which dictates which SM couplings are selected.
Denoting by $h_e$ ($h_q$) the helicity of the electron (quark) to which the
LQ couples (i.e.~$h_e~(h_q)=L$ or $R$, depending on the leptoquark), the
contribution to the cross section due to the interference terms is found to
be
$$\eqalign{
{1\over 4}\sum_{\rm spins} & \vert\Scm_{int}\vert^2 =
{}~{2048 \, \pi^3 \, k \, \alpha^3_{em} \over 2 p_5\cdot p_6 - M_{\small{S}}^2}
\sum_i c^e_{i{h_e}} c^q_{i{h_q}} { (2\, p_4\cdot p_6 - M_i^2) \over
(2\, p_4\cdot p_6 - M_i^2)^2 + M_i^2 \Gamma_i^2 }
\Bigl\{ {p_1\cdot p_4 \, p_5\cdot p_6 \over s} \cr
& - {1 \over 4 p_1\cdot p_5}
{1\over s} \bigl( s \,  \epsilon_k\cdot p_4 \, \epsilon_k\cdot p_5 \, (p_1
- p_5)\cdot p_6  + s \left[ p_1 \cdot p_6 \, p_4 \cdot p_5 - p_1 \cdot p_5
\, p_4 \cdot p_6 \right] \cr
& \qquad\qquad\qquad - s \, \epsilon_k\cdot
p_4 \, \epsilon_k\cdot p_6 \, p_1 \cdot p_5  + 2 \, p_1 \cdot p_4 \left[
p_1 \cdot p_5 \, p_2 \cdot p_6 - p_1 \cdot p_6 \, p_2 \cdot p_5 \right]
\bigr) \cr
& + {1 \over 2} Q_{\small{S}} {1 \over 2 p_2\cdot p_4 + M_{\small{S}}^2} \Bigl[
(\epsilon_k\cdot p_4)^2 p_5\cdot p_6 \cr
& \qquad\qquad\qquad\qquad  +
{p_2\cdot p_4 \over p_1\cdot p_5} \epsilon_k\cdot p_4 \left[
\epsilon_k\cdot p_6 \, p_1\cdot p_5 +  \epsilon_k\cdot p_5 \, (2p_5 -
p_1)\cdot p_6 \right] \Bigr] \cr
& - {1\over 4} (Q_{\small{S}}+1) {p_5\cdot p_6 \over
p_1\cdot p_4} \left[ 2 p_1\cdot p_4 - (\epsilon_k\cdot p_4)^2 \right] \cr
&
+ {1\over 4} (Q_{\small{S}}+1) {1 \over p_1\cdot p_4 \, p_1 \cdot p_5 } \bigl[
\epsilon_k\cdot p_4 \cdot \epsilon_k\cdot p_6 \,  p_1 \cdot p_5 \, ( p_4 -
p_1 ) \cdot p_2 \cr
& \qquad\qquad\qquad\quad + \epsilon_k\cdot p_4 \cdot
\epsilon_k\cdot p_5 \left( p_2 \cdot p_4 \,  p_5 \cdot p_6 + p_2\cdot (p_1
- p_4) \, p_6 \cdot (p_1-p_5) \right) \cr
& \qquad\qquad\qquad\quad +
p_1\cdot p_4 \left( p_1 \cdot p_5 \, p_2 \cdot p_6 - p_1 \cdot p_6 \,  p_2
\cdot p_5 \right) \cr
& \qquad\qquad\qquad\quad + s \, \left( p_1 \cdot p_6
\, p_4 \cdot p_5 - p_1 \cdot p_5 \, p_4 \cdot p_6 \right)/2  \bigr]
\Bigr\}, \cr}\eqno(8)
$$
where we have neglected the width $\Gamma_i$ in the numerator. In the above
equation, the standard model couplings to the electron and quark are given
by $c^e_{i{h_e}}$ and $c^q_{i{h_q}}$, respectively, in which $c_{iL}^f =
(v^f_i+a^f_i)/2$ and $c_{iR}^f = (v^f_i-a^f_i)/2$. As in Eq.~7, for those
terms in which two $\epsilon_k$'s appear, a sum is implicit over the two
photon polarizations.

The total cross section for $e\gamma\to e q{\overline q}$ is given by
integrating the sum of the three expressions in Eqs.~3,7 and 8 over the
allowed phase space. This must be done by Monte Carlo. It is clear that if
no cuts are applied, the standard model contribution dominates by many
orders of magnitude due to the fact that the gauge bosons can be on shell.
We have therefore applied stringent cuts in the integration in order to
considerably reduce the SM background. We have again considered two
collider energies: $\sqrt{s}=500$ GeV and 1 TeV. For the 500 GeV machine,
we required that the $p_{\small{T}}$ of the electron be greater than 100
GeV, that the invariant mass of the $q{\overline q}$ be greater than 100
GeV, and that the angle between the two jets satisfy
$\cos\theta_{q\overline q} < -0.4$. For the 1 TeV machine, we used
$p_{\small{T}}(e^-)>200$ GeV, $m_{q\overline q}>100$ GeV and
$\cos\theta_{q\overline q} < -0.2$. We also considered 3 different values
of $k$, the LQ coupling strength, at both values of $\sqrt{s}$.

Our results, for all LQ charges and chiralities, are shown in Figs.~5 and
6. Figs.~5a-d show the cross section at $\sqrt{s}=500$ GeV, for the 4
leptoquark charges, for $k=1,5,10$. For $k=1$, the cross section with both
LQ and SM contributions is essentially indistinguishable from that in which
only the SM is present. However, for the larger values of $k$, it is
possible to see significant deviations from the SM predictions (recall that
the expected integrated luminosity is 10 ${\rm fb}^{-1}$). Depending on the
value of $k$, a leptoquark with a mass of 700 GeV or even greater may be
observed. Note also that the left- and right-handed LQ's tend to give very
similar cross sections.

The cross sections at $\sqrt{s}=1$ TeV are shown in Figs.~6a-d. Here we
have taken $k=1,3,5$. Again, the curves with $k=1$ are fairly similar to
the SM prediction, but this time it may be possible to see a leptoquark
with a mass somewhat greater than the machine energy for some of the
charges (especially $Q_{em}=-5/3$), given that the expected integrated
luminosity is 60 ${\rm fb}^{-1}$). For $k=3$ or 5, on the other hand,
leptoquarks with a mass considerably larger than the kinematic limit are
observable. As at $\sqrt{s}=500$ GeV, the cross sections with left- and
right-handed LQ's are not very different from one another.

We remind the reader that a light ($M_{\small{S}}\leapprox 1$ TeV) charge
$-1/3$ left-handed LQ must have $k<0.1$ due to low-energy phenomenological
considerations. However, from Figs.~5 and 6, we see that, for both
$\sqrt{s}=500$ GeV and 1 TeV, it is necessary that $k\geapprox 1$ in order
that the signal for a LQ with a mass larger than $\sqrt{s}$ be observable.
Therefore the signal for this particular leptoquark is already excluded.
For this reason we have not considered the additional process $e\gamma\to
\nu_e d{\overline u}$, which is possible with a virtual charge $-1/3$
left-handed LQ, and which also has a SM background. For $k\simeq 0.1$, the
LQ contribution to this process would be much smaller than that of the SM.

{}From Figs.~5 and 6, it is clear that a larger energy machine is preferable
if one wishes to observe signals of virtual leptoquarks. Not only can one
simply probe larger masses, but smaller values of $k$ are allowed. The
point is that the largest contribution to the SM background comes from that
region of phase space which is nearest to the point where the gauge bosons
are produced on shell. As the energy of the machine increases, harder cuts
can be imposed, thereby further reducing the SM background. This is the
reason that the signals for LQ's with $k>1$ are more striking at
$\sqrt{s}=1$ TeV (Fig.~6) than at 500 GeV (Fig.~5).

To conclude, we have considered the possibility for the detection of
leptoquarks at $e\gamma$ colliders with $\sqrt{s}=500$ GeV and 1 TeV. It
turns out that $e\gamma$ colliders are excellent hunting grounds for LQ's
since these particles can be produced singly at such machines. Real
leptoquarks can be observed with masses essentially up to the
centre-of-mass energy even for couplings as small as
$O(10^{-3})$-$O(10^{-2})\alpha_{em}$. It is possible to find evidence for
the existence of LQ's with masses greater than the kinematic limit by
looking for signals of virtual leptoquark production in the process
$e\gamma\to eq{\overline q}$, for that region of parameter space in which
the LQ coupling is stronger than $\alpha_{em}$.
\vfill\eject

\noindent
\centerline{\bf Acknowledgments}

We would like to thank G. B\'elanger, A. Djouadi, J. Hewett and P.
Langacker for helpful discussions. This work was supported in part by the
Natural Sciences and Engineering Research Council of Canada, and by FCAR,
Qu\'ebec.
\vskip1truecm

\centerline{\bf References}
\vskip0.5truecm
\noindent
[1] I.F. Ginzburg, G.L. Kotkin, V.G. Serbo and V.I. Telnov, Pis'ma ZhETF
{\bf 34} (1981) 514; Sov.\ Yad.\ Fiz.\ {\bf 38} (1983) 372; Nucl.\ Instr.\
Methods {\bf 205} (1983) 47; I.F. Ginzburg, G.L. Kotkin, S.L. Panfil, V.G.
Serbo and V.I. Telnov, Sov.\ Yad.\ Fiz.\ {\bf 38} (1983) 1021; Nucl.\
Instr.\ Methods {\bf 219} (1984) 5; \hfil\break
[2] J.L. Hewett and T.G. Rizzo, Phys.\ Rep.\ {\bf 183} (1989) 193.
\hfil\break
[3] W. Buchm\"uller and D. Wyler, \plb{177}{86}{377}. \hfil\break
[4] Constraints on the parameter space of the LQ's with $Q_{em}\ne -1/3$
are presently being studied -- G. B\'elanger, private communication.
\hfil\break
[5] D. Decamp et.\ al., ALEPH Collaboration, CERN Report CERN-PPE/91-149
(1991), (submitted to Phys.\ Rep.); P. Abreu et.\ al., DELPHI
Collaboration, CERN Report CERN-PPE/91-138 (1991); B. Adeva et.\ al., L3
Collaboration, \plb{261}{91}{169}; G. Alexander et.\ al., OPAL
Collaboration, \plb{263}{91}{123}. \hfil\break
[6] T.G. Rizzo, \prd{44}{91}{186}. \hfil\break
[7] J. Alitti et.\ al., UA2 Collaboration, CERN Report CERN-PPE/91-158.
\hfil\break
[8] W. Buchm\"uller, R. R\"uckl and D. Wyler, \plb{191}{87}{442}.
\hfil\break
[9] V. Angelopoulos et.\ al., \npb{292}{87}{59}; J.F. Gunion and E. Ma,
\plb{195}{87}{257}. \hfil\break
[10] N. Harnew, in {\it Proceedings of the 1987 DESY Workshop on HERA
Physics}, Hamburg, Germany, October 1987. \hfil\break
[11] J.L. Hewett and S. Pakvasa, \plb{227}{89}{178}. \hfil\break

\vfill\eject
\centerline{Figure Captions}

\vskip1truecm
\noindent
1. Diagrams contributing to single leptoquark ($S$) production in $e\gamma$
collisions. The final state contains a $q$ if the LQ has fermion number
$F=0$ and a ${\overline q}$ if $F=2$.

\vskip0.6truecm
\noindent
2. Cross section for single leptoquark production in $e\gamma$ collisions
at (a) $\sqrt{s}=500$ GeV, (b) $\sqrt{s}=1$ TeV, for the 4 possible LQ
charges, $Q_{\small{S}}=-1/3,-2/3,-4/3,-5/3$. The results are given for
$k=1$.

\vskip0.6truecm
\noindent
3. Discovery region (25 events) for leptoquarks as a function of their mass
($M_{\small{S}}$) and coupling strength ($k$), in $e\gamma$ collisions at
(a) $\sqrt{s}=500$ GeV, (b) $\sqrt{s}=1$ TeV, for the 4 possible LQ
charges, $Q_{\small{S}}=-1/3,-2/3,-4/3,-5/3$. For a given curve, the
parameter space above the curve is observable.

\vskip0.6truecm
\noindent
4. Diagrams for the process $e\gamma\to eq{\overline q}$. The LQ diagrams
are (a),(b),(c); the SM diagrams are (d),(e). If the virtual LQ has fermion
number $F=0$, then the $q/{\overline q}$ assignments are those given in
parentheses; if $F=2$, they are not in parentheses. The momentum
assignments for all 5 diagrams are given in (a). Note that, regardless of
$F$, $p_4$ is assigned to the line which is connected to the initial $e$.

\vskip0.6truecm
\noindent
5. Total cross section for $e\gamma\to eq{\overline q}$ at $\sqrt{s}=500$
GeV, including both LQ and SM contributions, for (a) $Q_{\small{S}}=-1/3$,
(b) $Q_{\small{S}}=-2/3$, (c) $Q_{\small{S}}=-4/3$, and (d)
$Q_{\small{S}}=-5/3$, for both left-handed ($L$) and right-handed ($R$)
leptoquarks. The solid line has $k=1$, the dashed line has $k=5$, and the
dash-dot line has $k=10$. The SM prediction is the dotted line, and is 0.42
fb in (a),(d) and 0.43 fb in (b),(c).

\vskip0.6truecm
\noindent
6. Total cross section for $e\gamma\to eq{\overline q}$ at $\sqrt{s}=1$
TeV, including both LQ and SM contributions, for (a) $Q_{\small{S}}=-1/3$,
(b) $Q_{\small{S}}=-2/3$, (c) $Q_{\small{S}}=-4/3$, and (d)
$Q_{\small{S}}=-5/3$, for both left-handed ($L$) and right-handed ($R$)
leptoquarks. The solid line has $k=1$, the dashed line has $k=3$, and the
dash-dot line has $k=5$. The SM prediction is the dotted line, and is 0.13
fb in (a),(d) and 0.11 fb in (b),(c).

\bye